\documentstyle[aps,prl,preprint]{revtex}
\begin{document}
\draft
\author{W.-J. Huang and S.-C. Gou}
\address{Department of Physics,\\
National Changhua University of Education,\\
Changhua 50058, Taiwan}
\title{Ground state energy of the $f=1$ spinor\\
Bose-Einstein condensates }
\date{September 4, 1998}
\maketitle

\begin{abstract}
We calculate, in the standard Bogoliubov approximation, the ground state
energy of the spinor BEC with hyperfine spin $f=1$ where the two-body
repulsive hard-core and spin exchange interactions are both included. The
coupling constants characterized these two competing interactions are
expressed in terms of the corresponding $s$-wave scattering lengths using
second-order perturbation methods. We show that the ultraviolet divergence
arising in the ground state energy corrections can be exactly eliminated.
\end{abstract}


Recently, Stamper-Kurn {\it et} {\it al.}\cite{Stamper} have successfully
cooled $^{23}$Na atoms using an optical dipole trap and achieved BEC. In
their experiment, a multi-component BEC which is characterized by the three
hyperfine spin states $\left| f=1,m_{f}=\pm 1,0\right\rangle $ has been
observed. This has opened an interesting possibility to explore the
multi-component BEC with complicated internal spin dynamics in which not
only the global $U\left( 1\right) $ symmetry but also the rotational $%
SO\left( 3\right) $ symmetry in spin space are involved\cite{Ho,Ohmi}.

An important feature of the spinor condensate is that, in addition to the
repulsive binary hard-core collisions which give rise to the density-density
interaction, atoms in the condensates can also couple to each other via the
spin exchange interaction. Assuming that the interaction for each spin
exchange channel is again characterized by zero-range delta-potential
scattering, one thus obtains the interacting term ${\bf \hat{S}}\cdot {\bf 
\hat{S}}$ where ${\bf \hat{S}}$ is the spin density operator. The
competition between these two interactions thus lead to an intriguing
scenario of the spin dynamics of the spinor BEC which is characterized by a
complex ground state structure \cite{Ho,Ohmi,Law}.

The question arise now that how these two-body interactions alter the
dynamical properties of the condensates. It is known that in a weakly
interacting Bose condensed system, the two-body interactions play a crucial
role in determine the low temperature properties of the systems, which will
modify the ground state of the many-particle systems and cause a depletion
of the condensate fraction even at the zero temperature. Moreover, an
divergence could possibly appear when we calculate the ground state energy
in the standard Bogoliubov approximation. This divergence is due to the
naive assumption of a constant matrix element of binary interaction
irrespective of the relative momenta of the interacting particles. A well
illustrated example is the ultraviolet divergence occurring in the
calculated ground state energy of the one-component BEC where the two-body
interaction is described by the repulsive hard-core collisions with a
momentum-independent coupling constant\cite{Abrikosov}. To eliminate such a
divergence and gain more insights into the ground state properties of the
condensates, one has to calculate the $s$-wave scattering length at least to
the second order in coupling constant. By expanding the coupling constant in
powers of the $s$-wave scattering length the previously mentioned divergent
ground state energy can be rendered finite. This is expected since in a
physically sensible theory the ground state energy must assume a finite
value when expressed in terms of physically measurable quantities. In other
generalized Bose condensed systems such as the spinor BEC ultraviolet
divergences of the same sort could also appear. It is therefore quite
essential to verify if a similar procedure could completely removed these
divergences, and in this paper we address this issue in details for the $f=1$
spinor BEC in the presence of a constant magnetic field.

Consider an assembly of homogeneous dilute Bose gas with hyperfine spin $f=1$%
. The natural basis set to characterize such a system is the hyperfine spin
states $\left| m_{f}=\pm 1,0\right\rangle .$ However, in view of the special
symmetrical forms of the $S=1$ spin matrix representations , one may adopt
the basis set $\left\{ \left| x\right\rangle ,\left| y\right\rangle ,\left|
z\right\rangle \right\} $ which is defined as the eigenstates of the $\alpha 
$-th component of the spin operator with eigenvalue 0, i.e., $S_{\alpha
}\left| \alpha \right\rangle =0\left( \alpha =x,y,z\right) $ such that the
matrix elements for the $\alpha $-th spin component is given by $%
\left\langle \gamma \left| S_{\alpha }\right| \beta \right\rangle
=i\varepsilon _{\alpha \beta \gamma }$ where $\varepsilon _{\alpha \beta
\gamma }$ is the Levi-Civita tensor. This representation enables us to
relate the matrix elements of $S=1$ spin operators to those of the space
rotation, allowing the order parameter to behave as a vector under spin
space rotation.

For the $f=1$ spinor BEC, the bosonic atomic field can be described by the
multi-component field operator ${\bf \Psi },$ with components $\psi _{\alpha
}\left( {\bf r}\right) $ $\left( \alpha =x,y,z\right) $, and thus the
density of the particle number and spin can be written as $\hat{n}=\psi
_{\alpha }^{\dagger }\psi _{\alpha },$ and $\hat{S}_{\alpha }=\psi _{\beta
}^{\dagger }S_{\alpha }\psi _{\beta }=-i\epsilon _{\alpha \beta \gamma }\psi
_{\beta }^{\dagger }\psi _{\gamma }$ respectively. Note that we have used
the summation convention over the indices of component $\alpha ,\beta
,\cdots $ throughout this paper. Now, without loss of generality, the
Hamiltonian density can be constructed in the presence of a constant
magnetic field ${\bf B}$ pointing to the $z$-direction${\bf :}$ \cite{Ohmi} 
\begin{equation}
H=-\psi _{\alpha }^{\dagger }\frac{\nabla ^{2}}{2m}\psi _{\alpha }+\frac{1}{2%
}g_{n}\hat{n}^{2}+\frac{1}{2}g_{s}{\bf \hat{S}}\cdot {\bf \hat{S}-\Omega }%
\cdot {\bf \hat{S}\qquad }\left( \hbar =1\right)   \label{Ha}
\end{equation}
where ${\bf \Omega =}\Omega \hat{z}=g_{\mu }{\bf B}$ $\left( g_{\mu }:\text{%
gyromagnetic ratio}\right) $ is the Larmor frequency in a vectorial notation$%
.$ Expanding $\hat{n}$ and $\hat{S}_{\alpha }$ in terms of the field
operators, Eq.(\ref{Ha}) can be expressed as 
\begin{equation}
H=-\psi _{\alpha }^{\dagger }\frac{\nabla ^{2}}{2m}\psi _{\alpha }+\frac{1}{2%
}g_{1}\psi _{\beta }^{\dagger }\psi _{\alpha }^{\dagger }\psi _{\alpha }\psi
_{\beta }+\frac{1}{2}g_{2}\psi _{\beta }^{\dagger }\psi _{\beta }^{\dagger
}\psi _{\alpha }\psi _{\alpha }+i\epsilon _{\alpha \beta \gamma }\Omega
_{\gamma }\psi _{\alpha }^{\dagger }\psi _{\beta }  \label{Hamiltonian}
\end{equation}
where the two new coupling constants are given by $g_{1}=g_{n}+g_{s},$ $%
g_{2}=-g_{s}.$ According to the recent spectroscopic experiment by Abraham 
{\it et al}.\cite{Abraham}, it is conceivable in general that $g_{2}$ is
comparable to $g_{1}$ in magnitude and can be either positive or negative.
It is known that the positive $g_{2}$ implies the ferromagnetic coupling
while the negative one implies the antiferromagnetic coupling for the spin
exchange interaction.

Since the system is homogeneous, the field operator can be expanded in terms
of creation and annihilation operators characterized by momentum ${\bf k}$ 
\begin{equation}
\psi _{\alpha }\left( {\bf r}\right) =\frac{1}{\sqrt{V}}\sum_{{\bf k}%
}a_{\alpha ,{\bf k}}e^{i{\bf k}\cdot {\bf r}},
\end{equation}
where $V$ denotes the volume of the system. Accordingly, the Hamiltonian in
momentum space now reads as 
\begin{equation}
H=H_{0}+H_{\text{mag}}+H_{\text{int}}
\end{equation}
where 
\begin{equation}
H_{0}=\sum_{{\bf k}}\epsilon _{{\bf k}}a_{\alpha ,{\bf k}}^{\dagger
}a_{\alpha ,{\bf k}}  \label{HH0}
\end{equation}
\begin{equation}
H_{\text{mag}}=\sum_{{\bf k}}i\varepsilon _{\alpha \beta \gamma }\Omega
_{\gamma }a_{\alpha ,{\bf k}}^{\dagger }a_{\beta ,{\bf k}}  \label{HH1}
\end{equation}
\begin{eqnarray}
H_{\text{int}} &=&\frac{g_{1}}{2V}\sum_{{\bf k}_{1}+{\bf k}_{2}={\bf k}_{3}+%
{\bf k}_{4}}a_{\beta ,{\bf k}_{4}}^{\dagger }a_{\alpha ,{\bf k}%
_{3}}^{\dagger }a_{\alpha ,{\bf k}_{2}}a_{\beta ,{\bf k}_{1}}  \nonumber \\
&&+\frac{g_{2}}{2V}\sum_{{\bf k}_{1}+{\bf k}_{2}={\bf k}_{3}+{\bf k}%
_{4}}a_{\beta ,{\bf k}_{4}}^{\dagger }a_{\beta ,{\bf k}_{3}}^{\dagger
}a_{\alpha ,{\bf k}_{2}}a_{\alpha ,{\bf k}_{1}}  \label{HH2}
\end{eqnarray}
In the ground state, most particles occupy the ${\bf k}=0$ states. As a
result, the scattering between two nonzero-momentum states can be ignored
and the interacting part of the Hamiltonian can be replaced by 
\begin{eqnarray}
H_{\text{int}} &\simeq &\frac{g_{1}}{2V}\left[ a_{\beta ,0}^{\dagger
}a_{\alpha ,0}^{\dagger }a_{\alpha ,0}a_{\beta ,0}+\sum_{{\bf k}\neq
0}\left( a_{\beta ,{\bf k}}^{\dagger }a_{\alpha ,-{\bf k}}^{\dagger
}a_{\alpha ,0}a_{\beta ,0}+a_{\beta ,0}^{\dagger }a_{\alpha ,0}^{\dagger
}a_{\alpha ,{\bf k}}a_{\beta ,-{\bf k}}\right. \right.  \nonumber \\
&&\left. \left. +2a_{\beta ,{\bf k}}^{\dagger }a_{\alpha ,0}^{\dagger
}a_{\alpha ,{\bf k}}a_{\beta ,0}+2a_{\beta ,0}^{\dagger }a_{\alpha ,{\bf k}%
}^{\dagger }a_{\alpha ,{\bf k}}a_{\beta ,0}\right) \right]  \nonumber \\
&&+\frac{g_{2}}{2V}\left[ a_{\beta ,0}^{\dagger }a_{\beta ,0}^{\dagger
}a_{\alpha ,0}a_{\alpha ,0}+\sum_{{\bf k}\neq 0}\left( a_{\beta ,{\bf k}%
}^{\dagger }a_{\beta ,-{\bf k}}^{\dagger }a_{\alpha ,0}a_{\alpha
,0}+a_{\beta ,0}^{\dagger }a_{\beta ,0}^{\dagger }a_{\alpha ,{\bf k}%
}a_{\alpha ,-{\bf k}}\right. \right.  \nonumber \\
&&\left. \left. +4a_{\beta ,{\bf k}}^{\dagger }a_{\beta ,0}^{\dagger
}a_{\alpha ,{\bf k}}a_{\alpha ,0}\right) \right] .  \label{HH3}
\end{eqnarray}

Before calculating the $\ s$-wave scattering lengths to the second order, a
couple of remarks are in orders. First of all, for the sake of simplicity we
shall disregard the magnetic interaction $H_{\text{mag}}$ for a moment.
Secondly, the $s$-wave scattering lengths are formally determined from the
so-called $T$(ransition)-matrix which can be computed perturbatively by
using the diagrammatic techniques. Moreover, it is known that the energy
correction due to the two-body interactions can be directly related to the
matrix elements of the $T$-matrix \cite{Stoof}. Hence, one expects that the
desired $s$-wave scattering lengths can be obtained from the calculations of
energy corrections. In fact, it is not hard to show that, to the second
order, our results agree with those obtained by the $T$-matrix approach.
However, as the standard second-order perturbation methods are only required
in our paper, the calculations can be greatly simplified. With these remarks
in mind we are motivated to compute the energy corrections due to $H_{\text{%
int}}$.

We first introduce a class of 2-particle states defined by 
\begin{equation}
\left| 0,0;\varphi \right\rangle =\frac{1}{\sqrt{2}}\varphi _{\alpha
}^{*}\varphi _{\beta }^{*}a_{\beta ,0}^{\dagger }a_{\alpha ,0}^{\dagger
}\left| \text{vac}\right\rangle  \label{2-particle-state}
\end{equation}
where $\varphi _{\alpha }$ are constant parameters and $\left| \text{vac}%
\right\rangle $ is the Fock vacuum. Such states can be normalized by
imposing the condition $\varphi _{\alpha }^{*}\varphi _{\alpha }=\left|
\varphi \right| ^{2}=1.$ Quite clearly, the unperturbed energy vanishes in
the presence of the state Eq.(\ref{2-particle-state})$.$ It is easy to show
that the first-order energy corrections due to $H_{\text{int}}$ is 
\begin{equation}
E_{\text{int}}^{\left( 1\right) }=\left\langle 0,0;\varphi \left| H_{\text{%
int}}\right| 0,0;\varphi \right\rangle =\frac{g_{1}}{V}\left| \varphi
\right| ^{4}+\frac{g_{2}}{V}\left| \varphi ^{2}\right| ^{2},
\label{1st-E-int}
\end{equation}

Next, we consider the second-order correction for the energy. Now, in view
of the explicit form of $H_{\text{int}}$ given in Eq.(\ref{HH3})$,$ the only
possible intermediate states are those states of two non-condensate
particles carrying opposite momenta 
\begin{equation}
\left| {\bf k},-{\bf k};\alpha ,\beta \right\rangle \equiv a_{\alpha ,{\bf k}%
}^{\dagger }a_{\beta ,-{\bf k}}^{\dagger }\left| \text{vac}\right\rangle 
\end{equation}
with which the unperturbed energy is given by 
\begin{equation}
\left\langle {\bf k},-{\bf k};\alpha ,\beta \left| H_{0}\right| {\bf k},-%
{\bf k};\alpha ,\beta \right\rangle =\epsilon _{{\bf k}}+\epsilon _{-{\bf k}%
}=2\epsilon _{{\bf k}}.
\end{equation}
Thus the second-order energy correction due to $H_{\text{int}}$ is 
\begin{equation}
E_{\text{int}}^{\left( 2\right) }=-\frac{1}{2}\sum_{{\bf k}\neq 0}\frac{%
\left| \left\langle 0,0;\varphi \left| H_{\text{int}}\right| {\bf k},-{\bf k}%
;\alpha ,\beta \right\rangle \right| ^{2}}{2\epsilon _{{\bf k}}-0}.
\label{2nd-E-int}
\end{equation}
The factor $1/2$ in Eq.(\ref{2nd-E-int}) is inserted in order to avoid the
double counting of the momentum states. Now, we have 
\begin{eqnarray}
&&\left\langle 0,0;\varphi \left| H_{\text{int}}\right| {\bf k},-{\bf k}%
;\alpha ,\beta \right\rangle   \nonumber \\
&=&\frac{g_{1}}{V}\left\langle 0,0;\varphi \left| a_{\beta ,0}^{\dagger
}a_{\alpha ,0}^{\dagger }a_{\alpha ,{\bf k}}a_{\beta ,-{\bf k}}\right| {\bf k%
},-{\bf k};\alpha ,\beta \right\rangle   \nonumber \\
&&+\frac{g_{2}}{V}\left\langle 0,0;\varphi \left| a_{\beta ,0}^{\dagger
}a_{\beta ,0}^{\dagger }a_{\alpha ,{\bf k}}a_{\alpha ,-{\bf k}}\right| {\bf k%
},-{\bf k};\alpha ,\beta \right\rangle \delta _{\alpha \beta } \\
&=&\frac{\sqrt{2}}{V}\left[ g_{1}\varphi _{\alpha }\varphi _{\beta
}+g_{2}\varphi ^{2}\delta _{\alpha \beta }\right] ,  \nonumber
\end{eqnarray}
and hence 
\begin{eqnarray}
E_{\text{int}}^{\left( 2\right) } &=&-\frac{1}{V^{2}}\sum_{{\bf k}\neq 0}%
\frac{\left| g_{1}\varphi _{\alpha }\varphi _{\beta }+g_{2}\varphi
^{2}\delta _{\alpha \beta }\right| ^{2}}{2\epsilon _{{\bf k}}}  \nonumber \\
&=&-\frac{1}{V}\left[ g_{1}^{2}\left| \varphi \right| ^{4}+\left(
2g_{1}g_{2}+3g_{2}^{2}\right) \left| \varphi ^{2}\right| ^{2}\right] \int 
\frac{d^{3}k}{\left( 2\pi \right) ^{3}}\frac{1}{2\epsilon _{{\bf k}}}.
\label{Kop}
\end{eqnarray}
Obviously, the integral in Eq.(\ref{Kop}) diverges as $\left| {\bf k}\right| 
{\bf \rightarrow }\infty $. Choosing $\varphi _{\alpha }$ in such a way that 
$\left| \varphi ^{2}\right| =0$. yields 
\begin{equation}
g_{1}=E_{\text{int}}^{\left( 1\right) }V,
\end{equation}
indicating that $g_{1}$ is proportional to the first-order energy correction
due to the two-particle interaction $H_{\text{int}}$. At this order, $g_{1}$
is related to the corresponding $s$-wave scattering length $a_{1}$ by 
\begin{equation}
g_{1}=\frac{4\pi a_{1}}{m}.
\end{equation}
Hence, to the second order, $a_{1}$ is related to $g_{1}$ by the following
equation 
\begin{eqnarray}
\frac{4\pi a_{1}}{m} &\equiv &\tilde{g}_{1}=\left( E_{\text{int}}^{\left(
1\right) }+E_{\text{int}}^{\left( 2\right) }\right) V  \nonumber \\
&=&g_{1}-g_{1}^{2}\int \frac{d^{3}k}{\left( 2\pi \right) ^{3}}\frac{1}{%
2\epsilon _{{\bf k}}}.  \label{g1-correction}
\end{eqnarray}
Here $\tilde{g}_{1}$ will be referred to as the corrected coupling constant
of $g_{1}$. Writing the original coupling $g_{1}$ in terms of the corrected
coupling $\tilde{g}_{1},$ we have at the same order 
\begin{equation}
g_{1}=\tilde{g}_{1}+\tilde{g}_{1}{}^{2}\int \frac{d^{3}k}{\left( 2\pi
\right) ^{3}}\frac{1}{2\epsilon _{{\bf k}}}  \label{xxx}
\end{equation}
which is equivalent to the results demonstrated in the one-component case 
\cite{Abrikosov}. Next, we consider the corrections of $g_{2}$. Unlike\ $%
g_{1}$, $g_{2}$ can not be isolated directly in the present formalism.
Instead, we shall consider the sum $g_{1}+g_{2}$ which is nothing but the
coupling $g_{n}$ associated with the density-density interaction as $%
g_{1}=g_{n}+g_{s},$ $g_{2}=-g_{s}$. To this end, we may take $\varphi ^{2}=1$
such that $g_{1}+g_{2}=E_{\text{int}}^{\left( 1\right) }V$ and hence 
\begin{eqnarray}
\tilde{g}_{1}+\tilde{g}_{2} &=&\left( E_{\text{int}}^{\left( 1\right) }+E_{%
\text{int}}^{\left( 2\right) }\right) V  \nonumber \\
&=&g_{1}+g_{2}-\left( g_{1}^{2}+2g_{1}g_{2}+3g_{2}^{2}\right) \int \frac{%
d^{3}k}{\left( 2\pi \right) ^{3}}\frac{1}{2\epsilon _{{\bf k}}}
\label{gn-correction}
\end{eqnarray}
Subtracting Eq.(\ref{g1-correction}) from Eq.(\ref{gn-correction}) and using
Eq.(\ref{g1-correction}) again, we have at this order 
\begin{equation}
g_{2}=\tilde{g}_{2}+\left( 2\tilde{g}_{1}\tilde{g}_{2}+3\tilde{g}%
_{2}^{2}\right) \int \frac{d^{3}k}{\left( 2\pi \right) ^{3}}\frac{1}{%
2\epsilon _{{\bf k}}}  \label{g2-correction}
\end{equation}
Alternatively, $g_{n}$ and $g_{s}$ are related to the corresponding
corrected couplings by 
\begin{eqnarray}
g_{n} &=&\tilde{g}_{n}+\left( \tilde{g}_{n}^{2}+2\tilde{g}_{s}^{2}\right)
\int \frac{d^{3}k}{\left( 2\pi \right) ^{3}}\frac{1}{2\epsilon _{{\bf k}}}
\label{gsgn-correction} \\
g_{s} &=&\tilde{g}_{s}+\left( 2\tilde{g}_{n}\tilde{g}_{s}-\tilde{g}%
_{s}^{2}\right) \int \frac{d^{3}k}{\left( 2\pi \right) ^{3}}\frac{1}{%
2\epsilon _{{\bf k}}}  \nonumber
\end{eqnarray}
It should be noted that the corrected coupling constants, $\tilde{g}_{1}$
and $\tilde{g}_{2},$ are consistent with the one-loop corrections obtained
by using the Feynman diagram techniques \cite{Huang-Gou}. These results are
actually unaltered in the presence of a constant magnetic field. The point
is that the two-body interaction term $H_{\text{int}}$, in fact, commutes
with the magnetic term $H_{\text{mag}}$. As a consequence, despite that the
magnetic interaction would, inevitably, introduce a Zeeman energy shift to
each hyperfine spin state the total Zeeman energy is conserved in the
two-particle scattering processes. Based on this point, one can easily check
that both Eqs.(\ref{xxx}) and (\ref{g2-correction}) remain correct.

We now proceed to calculate the ground state energy with the foregoing
results. In the standard Bogoliubov approximation the operators $a_{\alpha
,0}$ and $a_{\alpha ,0}^{\dagger }$ are replaced by the classical number $%
\Phi _{\alpha }\sqrt{V}$ and $\Phi _{\alpha }^{*}\sqrt{V}$ respectively,
such that $\left| {\bf \Phi }\right| ^{2}=N_{0}/V=n_{0}$ represents the
density of condensate particles. Making these replacements into Eqs.(\ref
{HH1}) and (\ref{HH3}) yields 
\begin{equation}
H_{\text{mag}}\rightarrow iV\varepsilon _{\alpha \beta \gamma }\Omega
_{\gamma }\Phi _{\alpha }^{*}\Phi _{\beta }+\sum_{{\bf k}\neq 0}i\varepsilon
_{\alpha \beta \gamma }\Omega _{\gamma }a_{\alpha ,{\bf k}}^{\dagger
}a_{\beta ,{\bf k}}  \label{J1}
\end{equation}
and 
\begin{eqnarray}
H_{\text{int}} &\rightarrow &\frac{1}{2}g_{1}V\left| {\bf \Phi }\right| ^{4}+%
\frac{1}{2}g_{2}V\left| {\bf \Phi }^{2}\right| ^{2}  \nonumber \\
&&+\frac{g_{1}}{2}\sum_{{\bf k}\neq 0}\left( \Phi _{\alpha }\Phi _{\beta
}a_{\beta ,{\bf k}}^{\dagger }a_{\alpha ,-{\bf k}}^{\dagger }+\Phi _{\alpha
}^{*}\Phi _{\beta }^{*}a_{\alpha ,{\bf k}}a_{\beta ,-{\bf k}}\right. 
\nonumber \\
&&\left. +2\Phi _{\alpha }^{*}\Phi _{\beta }a_{\beta ,{\bf k}}^{\dagger
}a_{\alpha ,{\bf k}}+2\left| \Phi \right| ^{2}a_{\alpha ,{\bf k}}^{\dagger
}a_{\alpha ,{\bf k}}\right)  \nonumber \\
&&+\frac{g_{2}}{2}\sum_{{\bf k}\neq 0}\left( \Phi ^{2}a_{\beta ,{\bf k}%
}^{\dagger }a_{\beta ,-{\bf k}}^{\dagger }+\Phi ^{*2}a_{\alpha ,{\bf k}%
}a_{\alpha ,-{\bf k}}\right.  \nonumber \\
&&\left. +4\Phi _{\beta }^{*}\Phi _{\alpha }a_{\beta ,{\bf k}}^{\dagger
}a_{\alpha ,{\bf k}}\right)  \label{J2}
\end{eqnarray}

Using Eqs.(\ref{J1}) and (\ref{J2}) we obtain the effective Hamiltonian 
\begin{equation}
H_{\text{eff}}=H_{\text{con}}+H_{\text{non}}  \label{K0}
\end{equation}
where 
\begin{equation}
H_{\text{con}}=%
%TCIMACRO{\dint }
%BeginExpansion
\displaystyle \int %
%EndExpansion
d^{3}r\left[ i\epsilon _{\alpha \beta \gamma }\Omega _{\gamma }\Phi _{\alpha
}^{*}\Phi _{\beta }+\frac{g_{1}}{2}\left| {\bf \Phi }\right| ^{4}\allowbreak
+\frac{g_{2}}{2}\left| {\bf \Phi }^{2}\right| ^{2}\right]  \label{H-con}
\end{equation}
\begin{equation}
H_{\text{non}}=\sum_{{\bf k}\neq 0}\left( a_{\alpha ,{\bf k}}^{\dagger }%
{\cal L}_{\alpha \beta }a_{\beta ,{\bf k}}+\frac{1}{2}{\cal M}_{\alpha \beta
}^{*}a_{\alpha ,{\bf k}}a_{\beta ,-{\bf k}}+\frac{1}{2}{\cal M}_{\alpha
\beta }a_{\alpha ,{\bf k}}^{\dagger }a_{\beta ,-{\bf k}}^{\dagger }\right)
\label{H-non}
\end{equation}
are the Hamiltonians for the condensate and non-condensate part,
respectively. Here the matrix elements are given by 
\begin{eqnarray}
{\cal L}_{\alpha \beta } &=&\epsilon _{{\bf k}}\delta _{\alpha \beta
}+i\varepsilon _{\alpha \beta \gamma }\Omega _{\gamma }+g_{1}\left| {\bf %
\Phi }\right| ^{2}\delta _{\alpha \beta }+g_{1}\Phi _{\beta }^{*}\Phi
_{\alpha }+2g_{2}\Phi _{\alpha }^{*}\Phi _{\beta }  \label{Lab} \\
{\cal M}_{\alpha \beta } &=&g_{1}\Phi _{\alpha }\Phi _{\beta }+g_{2}\Phi
^{2}\delta _{\alpha \beta }.  \nonumber
\end{eqnarray}
Note that $H_{\text{eff}}$ is precisely the Hartree-Fock-Bogoliubov
Hamiltonian in the standard Bogoliubov approximation\cite{Huang-Gou} whose
ground state structure can be determined by minimizing the integrand in Eq.(%
\ref{H-con}). As a result, two different ground state structures are found 
\cite{Ho,Ohmi}: 
\begin{equation}
{\bf \Phi =}\sqrt{n_{0}}\left( 1/\sqrt{2},i/\sqrt{2},0\right) \text{ for }%
n_{0}g_{2}>-\Omega  \label{ferro}
\end{equation}
which is referred to as the ``ferromagnetic'' state and 
\begin{equation}
{\bf \Phi =}\sqrt{n_{0}}\left( \cos \theta ,i\sin \theta ,0\right) \text{
for }n_{0}g_{2}<-\Omega  \label{polar}
\end{equation}
as the ``polar'' state. Here the cosine and the sine in Eq.(\ref{polar}) are
given by 
\begin{eqnarray}
\cos \theta &=&\frac{1}{2}\left( \sqrt{1+\frac{\Omega }{\left| g_{2}\right|
n_{0}}}+\sqrt{1-\frac{\Omega }{\left| g_{2}\right| n_{0}}}\right)  \nonumber
\\
\sin \theta &=&\frac{1}{2}\left( \sqrt{1+\frac{\Omega }{\left| g_{2}\right|
n_{0}}}-\sqrt{1-\frac{\Omega }{\left| g_{2}\right| n_{0}}}\right)
\label{costh}
\end{eqnarray}

Since $H_{\text{eff}}$ is quadratic in $a_{\alpha ,{\bf k}}^{\dagger }$ and $%
a_{\alpha ,{\bf k}}$, we can diagonalize this Hamiltonian by using the
generalized Bogoliubov transformation 
\begin{equation}
a_{\alpha ,{\bf k}}=\sum_{i}\left[ u_{\alpha ,{\bf k}}^{\left( i\right) }b_{%
{\bf k}}^{\left( i\right) }-v_{\alpha ,-{\bf k}}^{\left( i\right) }b_{-{\bf k%
}}^{\dagger \left( i\right) }\right]  \label{quasi-particle}
\end{equation}
where $i$ is the mode index and $\left| u_{\alpha ,{\bf k}}^{\left( i\right)
}\right| ^{2}-\left| v_{\alpha ,{\bf k}}^{\left( i\right) }\right| ^{2}=1$.
In terms of the quasiparticle creation and annihilation operators $b_{{\bf k}%
}^{\left( i\right) }$ and $b_{{\bf k}}^{\dagger \left( i\right) }$, the
Hamiltonian takes the following form 
\begin{equation}
H_{\text{eff}}=H_{\text{con}}+\sum_{i}\sum_{{\bf k}\neq 0}E_{{\bf k}%
}^{\left( i\right) }\left( b_{{\bf k}}^{\dagger \left( i\right) }b_{{\bf k}%
}^{\left( i\right) }-\left| v_{\alpha ,{\bf k}}^{\left( i\right) }\right|
^{2}\right)  \label{Heff}
\end{equation}
We now define the ground state which is annihilated by all $b_{{\bf k}%
}^{\left( i\right) }$ i.e., $b_{{\bf k}}^{\left( i\right) }\left| \text{GND}%
\right\rangle =0,$ such that the ground state energy is found to be 
\begin{equation}
E_{\text{GND}}=V\left[ i\varepsilon _{\alpha \beta \gamma }\Omega _{\gamma
}\Phi _{\alpha }^{*}\Phi _{\beta }+\frac{g_{1}}{2}\left| {\bf \Phi }\right|
^{4}\allowbreak +\frac{g_{2}}{2}\left| {\bf \Phi }^{2}\right| ^{2}-\int 
\frac{d^{3}k}{\left( 2\pi \right) ^{3}}\sum_{i}E_{{\bf k}}^{\left( i\right)
}\left| v_{\alpha ,{\bf k}}^{\left( i\right) }\right| ^{2}\right] ,
\label{E-GND}
\end{equation}
where the last integral indicates the energy shift due to the quasiparticle
excitations. To calculate the ground state energy, one needs to know
precisely the values of $E_{{\bf k}}^{\left( i\right) }$ and $v_{\alpha ,%
{\bf k}}^{\left( i\right) }$ for the quasiparticle modes. This can be done
by using the standard Hartree-Fock-Bogoliubov mean-field method, and we have
calculated $E_{{\bf k}}^{\left( i\right) }$ and $v_{\alpha ,{\bf k}}^{\left(
i\right) }$ for the quasiparticle modes. In the following, we devote our
attention to the case in which the ground state is ``polar'', since for the
``ferromagnetic'' case the results are identical the those of the
one-component scalar BEC \cite{Abrikosov} and can be obtained from the
``polar'' case by setting $\Omega =-n_{0}g_{2}$. As a result, we find that
the low lying excitations can be described by two gapless modes $E_{{\bf k}%
}^{\left( \pm \right) }$ and one massive mode $E_{{\bf k}}^{\left( 0\right)
} $: 
\begin{equation}
E_{{\bf k}}^{\left( \pm \right) }=\sqrt{\epsilon _{{\bf k}}\left( \epsilon _{%
{\bf k}}+2n_{0}g^{\left( \pm \right) }\right) },\qquad E_{{\bf k}}^{\left(
0\right) }=\sqrt{\epsilon _{{\bf k}}\left( \epsilon _{{\bf k}%
}+2n_{0}g^{\left( 0\right) }\right) +\Omega ^{2}},  \label{spectrum}
\end{equation}
for which the corresponding nonvanishing distribution functions are given by 
\cite{Huang-Gou} 
\begin{equation}
\left( 
\begin{array}{l}
v_{x,{\bf k}}^{\left( \pm \right) } \\ 
v_{y,{\bf k}}^{\left( \pm \right) }
\end{array}
\right) =\left( 
\begin{array}{l}
+A^{\left( \pm \right) } \\ 
-B^{\left( \pm \right) }
\end{array}
\right) \beta _{{\bf k}}^{\left( \pm \right) }\qquad v_{z,{\bf k}}^{\left(
0\right) }=-\left( 1-\frac{\Omega ^{2}}{n_{0}^{2}g_{2}^{2}}\right) \beta _{%
{\bf k}}^{\left( 0\right) }  \label{distribution}
\end{equation}
where 
\begin{equation}
g^{\left( \pm \right) }=\frac{1}{2}\left[ g_{1}\pm \sqrt{g_{1}^{2}+4g_{2}%
\left( g_{1}+g_{2}\right) \left( 1-\frac{\Omega ^{2}}{n_{0}^{2}g_{2}^{2}}%
\right) }\right] ,\qquad g^{\left( 0\right) }=\left| g_{2}\right|
\end{equation}
\begin{equation}
\beta _{{\bf k}}^{\left( i\right) }=\frac{n_{0}g^{\left( i\right) }}{\sqrt{%
2E_{{\bf k}}^{\left( i\right) }\left( E_{{\bf k}}^{\left( i\right)
}+\epsilon _{{\bf k}}+n_{0}g^{\left( i\right) }\right) }}\qquad \left( i=\pm
,0\right)
\end{equation}
\begin{equation}
A^{\left( \pm \right) }=\frac{g_{1}\Omega /\left| g_{2}\right| }{\sqrt{%
\left. \eta ^{\left( \pm \right) }\right. ^{2}+\left( g_{1}\Omega
/g_{2}\right) ^{2}}},B^{\left( \pm \right) }=\frac{-i\eta ^{\left( \pm
\right) }}{\sqrt{\left. \eta ^{\left( \pm \right) }\right. ^{2}+\left(
g_{1}\Omega /g_{2}\right) ^{2}}}  \label{AB-plus-minus}
\end{equation}
\begin{equation}
\eta ^{\left( \pm \right) }=n_{0}\left( g_{1}-2g^{\left( \pm \right)
}\right) +n_{0}\left( g_{1}+2g_{2}\right) \sqrt{1-\left( \Omega /n_{0}\left|
g_{2}\right| \right) ^{2}}  \label{39}
\end{equation}
Moreover, with the condensate wavefunctions described in Eqs.(\ref{polar})
and (\ref{costh}), we obtain the following results: 
\begin{equation}
\left| {\bf \Phi }\right| ^{2}=n_{0},\qquad \left| {\bf \Phi }^{2}\right|
^{2}=n_{0}^{2}\left( 1-\frac{\Omega ^{2}}{n_{0}^{2}g_{2}^{2}}\right) ,\qquad
i\varepsilon _{\alpha \beta \gamma }\Omega _{\gamma }\Phi _{\alpha }^{*}\Phi
_{\beta }=-\frac{\Omega }{\left| g_{2}\right| }  \label{sham}
\end{equation}
However, one see that all the three integrals 
\[
\int \frac{d^{3}k}{\left( 2\pi \right) ^{3}}E_{{\bf k}}^{\left( i\right)
}\left| v_{\alpha ,{\bf k}}^{\left( i\right) }\right| ^{2}\qquad \left(
i=\pm ,0\right) 
\]
are divergent when $\left| {\bf k}\right| \rightarrow \infty $. Note also
that they are essentially of second-order in the coupling constants $g_{1}$
and $g_{2}$. To eliminate these ultraviolet divergences we substitute Eqs.(%
\ref{xxx}) and (\ref{g2-correction}) into Eq.(\ref{E-GND}). The resulting
expression for the ground state energy is 
\begin{eqnarray}
E_{\text{GND}} &=&V\left[ i\varepsilon _{\alpha \beta \gamma }\omega
_{\gamma }\Phi _{\alpha }^{*}\Phi _{\beta }+\frac{\tilde{g}_{1}}{2}\left| 
{\bf \Phi }\right| ^{4}\allowbreak +\frac{\tilde{g}_{2}}{2}\left| {\bf \Phi }%
^{2}\right| ^{2}\right]  \nonumber \\
&&+V%
%TCIMACRO{\dint }
%BeginExpansion
\displaystyle \int %
%EndExpansion
\frac{d^{3}k}{\left( 2\pi \right) ^{3}}\frac{1}{2\epsilon _{{\bf k}}}\left\{
\left[ \frac{\tilde{g}_{1}^{2}}{2}\left| {\bf \Phi }\right| ^{4}+\frac{1}{2}%
\left( 2\tilde{g}_{1}\tilde{g}_{2}+3\tilde{g}_{2}^{2}\right) \left| {\bf %
\Phi }^{2}\right| ^{2}\right] \right.  \nonumber \\
&&\left. -\sum_{i}E_{{\bf k}}^{\left( i\right) }\left( \tilde{g}_{1},\tilde{g%
}_{2}\right) \left| v_{\alpha ,{\bf k}}^{\left( i\right) }\left( \tilde{g}%
_{1},\tilde{g}_{2}\right) \right| ^{2}\right\} .  \label{hope}
\end{eqnarray}
Note that the condensate wavefunction ${\bf \Phi }$ determined by minimizing
the sum of terms in the first line of Eq.(\ref{hope}) has the same form as
that in Eq.(\ref{sham}) except that the corrected coupling constants are
substituted instead. Furthermore, the last term in Eq.(\ref{hope}) can be
expanded in powers of the corrected coupling constants. Since Eq.(\ref{hope}%
) is valid only up to the second order in the corrected couplings, it
suffices to substitute $g_{1}=\tilde{g}_{1},g_{2}=\tilde{g}_{2}$ in the
expressions for $E_{{\bf k}}^{\left( i\right) }$ and $v_{\alpha ,{\bf k}%
}^{\left( i\right) },$ i.e., in Eqs.(\ref{spectrum})-(\ref{39}). On these
grounds, we are now ready to calculate $E_{\text{GND}}$ given by Eq.(\ref
{hope}). First, we note that 
\begin{equation}
\left. \tilde{g}^{\left( +\right) }\right. ^{2}+\left. \tilde{g}^{\left(
-\right) }\right. ^{2}=\tilde{g}_{1}^{2}+2\tilde{g}_{1}\left( \tilde{g}_{1}+%
\tilde{g}_{2}\right) \left( 1-\frac{\Omega ^{2}}{n_{0}^{2}\tilde{g}_{2}^{2}}%
\right) ,
\end{equation}
and hence 
\begin{eqnarray}
&&\tilde{g}_{1}\left| {\bf \Phi }\right| ^{4}+\left( 2\tilde{g}_{1}\tilde{g}%
_{2}+3\tilde{g}_{2}^{2}\right) \left| {\bf \Phi }^{2}\right| ^{2}  \nonumber
\\
&=&n_{0}^{2}\left[ \left. \tilde{g}^{\left( +\right) }\right. ^{2}+\left. 
\tilde{g}^{\left( -\right) }\right. ^{2}+\left( 1-\frac{\Omega ^{2}}{%
n_{0}^{2}\tilde{g}_{2}^{2}}\right) \left. \tilde{g}^{\left( 0\right)
}\right. ^{2}\right]
\end{eqnarray}
The integral in Eq.(\ref{hope}) is then equal to 
\begin{eqnarray}
&&\frac{1}{2}n_{0}^{2}V\int \frac{d^{3}k}{\left( 2\pi \right) ^{3}}\left[
\sum_{i=\pm }\left. \tilde{g}^{\left( i\right) }\right. ^{2}\left( \frac{1}{%
2\epsilon _{{\bf k}}}-\frac{1}{\left( E_{{\bf k}}^{\left( i\right)
}+\epsilon _{{\bf k}}+n_{0}\tilde{g}^{\left( i\right) }\right) }\right)
\right.  \nonumber \\
&&\left. +\left( \tilde{g}_{2}^{2}-\frac{\Omega ^{2}}{n_{0}^{2}}\right)
\left( \frac{1}{2\epsilon _{{\bf k}}}-\frac{1}{\left( E_{{\bf k}}^{\left(
0\right) }+\epsilon _{{\bf k}}+n_{0}\left| \tilde{g}_{2}\right| \right) }%
\right) \right] .
\end{eqnarray}
Note that the first two terms are the same as that of the one-component case 
\cite{Abrikosov} 
\begin{eqnarray}
&&\frac{1}{2}n_{0}^{2}V\int \frac{d^{3}k}{\left( 2\pi \right) ^{3}}\left. 
\tilde{g}^{\left( \pm \right) }\right. ^{2}\left( \frac{1}{2\epsilon _{{\bf k%
}}}-\frac{1}{\left( E_{{\bf k}}^{\left( \pm \right) }+\epsilon _{{\bf k}%
}+n_{0}\tilde{g}^{\left( \pm \right) }\right) }\right)  \nonumber \\
&=&\frac{2\pi n_{0}^{5/2}}{m}V\left( \frac{128}{15\sqrt{\pi }}\left.
a^{\left( \pm \right) }\right. ^{5/2}\right) \qquad
\end{eqnarray}
where $a^{\left( \pm \right) }=m\tilde{g}^{\left( \pm \right) }/4\pi $ are
the corresponding $s$-wave scattering wavelengths. The last integral can be
expressed as 
\begin{eqnarray}
&&\frac{1}{2}n_{0}^{2}V\left( \tilde{g}_{2}^{2}-\frac{\Omega ^{2}}{n_{0}^{2}}%
\right) \int \frac{d^{3}k}{\left( 2\pi \right) ^{3}}\frac{E_{{\bf k}%
}^{\left( 0\right) }-\epsilon _{{\bf k}}+n_{0}\left| \tilde{g}_{2}\right| }{%
2\epsilon _{{\bf k}}\left( E_{{\bf k}}^{\left( 0\right) }+\epsilon _{{\bf k}%
}+n_{0}\left| \tilde{g}_{2}\right| \right) }  \nonumber \\
&=&\frac{2\pi V}{m}n_{0}^{5/2}\left( \frac{128}{15\sqrt{\pi }}\left|
a_{2}\right| ^{5/2}\right) \left( 1-t^{2}\right) F\left( t^{2}\right)
\end{eqnarray}
where $t=\Omega /n_{0}\left| \tilde{g}_{2}\right| $, and the function of
integral is defined as 
\begin{equation}
F\left( t^{2}\right) =\frac{15\sqrt{2}}{32}\int_{0}^{\infty }dx\frac{1-x^{2}+%
\sqrt{t^{2}+2x^{2}+x^{4}}}{1+x^{2}+\sqrt{t^{2}+2x^{2}+x^{4}}}\quad \text{for 
}0\leq t^{2}\leq 1,  \label{F(a)}
\end{equation}
which is a monotonically increasing function that can not be analytically
evaluated in general.

Finally, using Eq.(\ref{sham}) we get 
\begin{eqnarray}
&&\left[ i\varepsilon _{\alpha \beta \gamma }\omega _{\gamma }\Phi _{\alpha
}^{*}\Phi _{\beta }+\frac{\tilde{g}_{1}}{2}\left| {\bf \Phi }\right|
^{4}\allowbreak +\frac{\tilde{g}_{2}}{2}\left| {\bf \Phi }^{2}\right|
^{2}\right] V  \nonumber \\
&=&\left[ \frac{\tilde{g}_{1}n_{0}^{2}}{2}+\frac{\tilde{g}_{2}n_{0}^{2}}{2}%
\left( 1-\frac{\Omega ^{2}}{n_{0}^{2}\tilde{g}_{2}^{2}}\right) -\frac{\Omega 
}{\left| \tilde{g}_{2}\right| }\right] V  \nonumber \\
&=&\frac{2\pi n_{0}^{2}V}{m}\left( a_{n}-t^{2}a_{s}\right)
\end{eqnarray}
and therefore the ground state energy is given by 
\begin{eqnarray}
E_{\text{GND}} &=&\frac{2\pi n_{0}^{2}V}{m}\left[ \left(
a_{n}-t^{2}a_{s}\right) +\frac{128}{15\sqrt{\pi }}n_{0}^{1/2}\left( \left.
a^{\left( +\right) }\right. ^{5/2}+\left. a^{\left( -\right) }\right.
^{5/2}\right. \right.  \nonumber \\
&&\left. \left. +\left( 1-t^{2}\right) F\left( t^{2}\right)
a_{s}^{5/2}\right) \right] ,  \label{100}
\end{eqnarray}
where $a_{n}=m\tilde{g}_{n}/4\pi ,a_{s}=m\tilde{g}_{s}/4\pi .$ The terms
proportional to $\left. a^{\left( \pm \right) }\right. ^{5/2}$ are caused by
the two gapless modes and have the same form of the phonon-like mode in the
one-component BEC. The last term in Eq.(\ref{100}) is due to the massive
mode, which depends solely on the scattering length $a_{s}$ for the spin
exchange channel and is suppressed by the increasing magnetic field.

In conclusion, we have analytically calculated the ground state energy of a
homogeneous spinor BEC with hyperfine spin $f=1$ based on the Bogoliubov
approximation$.$ In this weakly interacting system, the two-body
interactions are described by the hard-core collisions and the spin exchange
interaction which are characterized by the coupling constants $g_{1}$ and $%
g_{2}$ respectively. Using the second-order perturbation methods, the two
bare coupling constants $g_{1}$ and $g_{2},$ are expressed in terms of their
corrected ones, $\tilde{g}_{1}$ and $\tilde{g}_{2}$ which are directly
related to the physically measurable $s$-wave wavelengths for the
corresponding scattering channels. It is found that the correction of $g_{1}$
has the same form as that of the one-component scalar BEC. However, the
correction of $g_{2}$ is more complicated and has dependence on the
corrected coupling constant $\tilde{g}_{1}.$ With the corrected coupling
constants, we are able to show that the ultraviolet divergence occurring in
the calculation of ground state energy can be completely removed.

This work is supported by the National Science Council, Taiwan under Grant
NSC-88-2112--M-018-004.

\end{document}